\documentstyle[aps,psfig]{revtex}

\begin{document}

\nopagebreak
\title{\hspace{5.0in}Fermilab-Pub-98/287 \vspace{0.5in}
Collision Integrals and the Generalized Kinetic Equation for Charged 
Particle Beams}
\author{~Stephan~I.~Tzenov}
\address{{\it Fermi National Accelerator Laboratory }\\
{\it P.~O.~Box 500, Batavia, IL 60510, USA}\\
{\it E.~mail: tzenov@fnal.gov}}
\maketitle

\newlength{\figwid} \newlength{\figlen} \medskip 

\begin{abstract}

In the present paper we study the role of particle interactions on the 
evolution of a high energy beam. The interparticle forces taken into 
account are due to space charge alone. We derive the collision integral 
for a charged particle beam in the form of Balescu-Lenard and Landau and 
consider its further simplifications. Finally, the transition to the 
generalized kinetic equation has been accomplished by using  the method 
of adiabatic elimination of fast variables.

\end{abstract}

\section{Introduction.}

In most of the works so far, dedicated to the study of beam plasma
properties the effect of interparticle collisions has been neglected. In
many important cases this is a sensible approximation giving satisfactory
results, yet one has to elucidate the limits of validity of ''collisionless
beam'' approach and to investigate the role of collision phenomena in beam
physics. Collisions are expected to bring about effects such as
thermalization, resistivity, diffusion etc. that influence the long term
behaviour of charged particle beams. The reasoning commonly adopted for
employing the ''collisionless beam'' approach is that characteristic
beam-plasma frequencies are much greater than collision frequencies for a
large number of situations in beam physics. Such an assumption is not based
on stable physical grounds as pointed in \cite{KlimU}.

The term ''collisionless beam'' means that interactions between particles
giving rise to dissipation and hence leading to establishment of equilibrium
state are not taken into account. In a number of cases involving reasonable
approximations it is sufficient to compute the macroscopic characteristics
(charge and current densities) in relatively big volume elements containing
a large number of particles. As a result interaction manifests itself in the
form of a mean, self-consistent field thus preserving the reversible
character of the dynamics involved, and leading to the time \textit{%
reversible} Vlasov's equation.

The notion of ''collisional beam'' usually conceived as the counterpart of
''collisionless beam'' implies that dissipation due to redistribution of
beam particles is taken into account, resulting in additional term (in the
form of Landau or Balescu-Lenard) in the kinetic equation. In a sense,
Landau and Vlasov approximations correspond to two limit cases: namely the
Landau collision integral takes into account interactions that determine
dissipation while the effect of the mean, self-consistent field is not
included into the physical picture involved. On the contrary, the latter is
the only way interactions manifest themselves in the Vlasov equation,
leaving however the question about the role of collisions near particle-wave
resonances unanswered. The Balescu-Lenard approximation lies somewhat in
between Landau and Vlasov limit cases with the due account of dynamic
polarization of the beam, that is a more complete inclusion of collective
effects resulting from interactions between charged particles.

In the present paper we derive the collision integrals for charged particle
beams. The transition to the unified kinetic, hydrodynamic and diffusion
description of particle beam propagation embedded in the generalized kinetic
equation \cite{TzArc} is further accomplished building on the concept of a
coarse-grained hydrodynamic picture. The latter implies the existence (and
their proper definition) of characteristic spacial and temporal scales typical 
for the hydrodynamic level of description \cite{KlimU}, \cite{KlimS}. Within 
the elementary cell of continuous medium thus defined it is naturally assumed
that local equilibrium state is reached. This state is further described
(defining the drift and diffusion coefficients in coordinate space) by the
method of adiabatic elimination of fast variables, widely used to match the
transition to Smoluchowski equation \cite{Gardiner}. The granulation of
phase space with the due account of concrete structure of continuous medium
results in additional collision integral in the kinetic equation, thus 
describing the dissipation caused by spacial diffusion of the distribution 
function and redistribution of particle coordinates \cite{TzArc}, \cite{KlimS}.

The generalized kinetic equation makes it possible to build an unified
picture of non equilibrium processes on kinetic and hydrodynamic scales
without involving a perturbation expansion in Knudsen number \cite{KlimS}%
. It can be shown that the set of hydrodynamic equations for cold beams put
in appropriate form is equivalent to mesoscopic quantum-like description of
particle beam propagation \cite{TzHyd}, \cite{TzSyn}.

The scope of the presentation given in the paper is as follows. In Sections
II and III we formulate and solve the equation for the fluctuations of the
microscopic phase space density in the case of a space charge dominated high
energy beam. The solution obtained provides the grounds to find explicitly
the collision integral in the form of Balescu-Lenard. Sections IV - VI deal
with the various forms and simplifications of the collision integral. In
Section VII we derive the additional Fokker-Planck term in the generalized
kinetic equation. Finally, Section VIII presents the conclusions of our
study.

\section{Averaged Microscopic Equations.}

In a previous paper \cite{TzArc} we derived the equation for the microscopic
phase space density with a small source, taking into account the proper
physical definition of continuous medium. It was the starting point in the
transition to the generalized kinetic equation for the one-particle
distribution function. The equation for the microscopic phase space density
reads as

$$
\frac{\partial N}{\partial \theta }+R\left( \widehat{\mathbf{v}}\cdot 
\widehat{\mathbf{\nabla }}_x\right) N+R\left[ \widehat{\mathbf{\nabla }}%
_p\cdot \left( \widehat{\mathbf{F}}_0+\widehat{\mathbf{F}}^{\left( M\right)
}\right) \right] N=\frac 1{\theta _{ph}}\left( \widehat{N}-N\right), 
\eqno (2.1)
$$

\noindent where $N=N\left( \widehat{\mathbf{x}},\widehat{\mathbf{p}}^{\left(
k\right) };\theta \right) $ is the true microscopic phase space density
written in the variables

$$
\widehat{\mathbf{x}}=\left( \widehat{x},\widehat{z},\widehat{\sigma }\right)
\qquad ;\qquad \widehat{\mathbf{p}}^{\left( k\right) }=\left( \widehat{p}%
_x^{\left( k\right) },\widehat{p}_z^{\left( k\right) },\widehat{\eta }%
^{\left( k\right) }\right) .  \eqno (2.2)
$$

\noindent They are related to the canonical coordinates ${\mathbf{x}} = \left(
x,z,\sigma \right) $ and canonical momenta ${\mathbf{p}} = \left(
p_x,p_z,h\right) $ through the following equations

$$
{\widehat{u}}=u-\eta {\mathcal{D}}_u\quad ;\quad {\widehat{p}}_u^{\left( k\right)
}={\widetilde{p}}_u^{\left( k\right) }-\eta \frac{d{\mathcal{D}}_u}{ds}\quad
;\quad {\widehat{\sigma }}=\sigma +\sum_{u=\left( x,z\right)} \left( u \frac{d%
{\mathcal{D}}_u}{ds}-{\widetilde{p}}_u {\mathcal{D}}_u\right) ,  \eqno (2.3a)
$$

$$
\widehat{\eta }^{\left( k\right) }=h^{\left( k\right) }-\frac 1{\beta
_o^2}\quad ;\quad \eta =h-\frac 1{\beta _o^2}\quad ;\quad h=\frac{\mathcal{H}%
}{\beta _o^2E_o},  \eqno (2.3b)
$$

$$
p_u^{\left( k\right) }=p_u-qA_u\qquad ;\qquad h^{\left( k\right) }=h-\frac{%
q\varphi }{\beta _o^2E_o},  \eqno (2.3c)
$$

\noindent where $u=\left( x,z\right) $ and all other notations are the same
as in Ref. \cite{TzArc}. In particular the following designations

$$
\widehat{\mathbf{v}}=\left( \widehat{p}_x^{\left( k\right) },\widehat{p}%
_z^{\left( k\right) },-{\mathcal{K}}{\widehat{\eta}}^{\left( k\right) }\right)
\qquad ;\qquad \widehat{\mathbf{\nabla }}_x=\left( \frac \partial {\partial 
\widehat{x}},\frac \partial {\partial \widehat{z}},\frac \partial {\partial 
\widehat{\sigma }}\right) ,  \eqno (2.4a)
$$

$$
\widehat{\mathbf{\nabla }}_p=\left( \frac \partial {\partial \widehat{p}%
_x^{\left( k\right) }},\frac \partial {\partial \widehat{p}_z^{\left(
k\right) }},\frac \partial {\partial \widehat{\eta }^{\left( k\right)
}}\right) ,  \eqno (2.4b)
$$

$$
\widehat{\mathbf{F}}_0=\left( -\frac{\partial \mathcal{U}}{\partial \widehat{%
x}},-\frac{\partial \mathcal{U}}{\partial \widehat{z}},\frac 1{2\pi R}\frac{%
\Delta E_0}{\beta _o^2E_o}\sin \left( \frac{\omega \widehat{\sigma }}{c\beta
_o}+\Phi _0\right) \right) ,  \eqno (2.4c)
$$

\[
\widehat{\mathbf{F}}^{\left( M\right) }=\frac q{\beta _o^2E_o}\left\{ \left(
1+\widehat{\mathbf{x}}\cdot \mathbf{K}\right) \left[ {\mathbf{E}}^{\left(
M\right) }+v_o\left( {\mathbf{e}}_s\times {\mathbf{B}}^{\left( M\right) }\right)
\right] +{\mathbf{e}}_s\left( \widehat{\mathbf{p}}^{\left( k\right) }\cdot 
\mathbf{E}^{\left( M\right) }\right) \right\} + 
\]

$$
+\frac q{p_o}\left( \widehat{\mathbf{p}}^{\left( k\right) }\times \mathbf{B}%
^{\left( M\right) }\right) _u,  \eqno (2.4d)
$$

$$
{\mathbf{e}}_s=\left( 0,0,1\right)  \eqno (2.4e)
$$

\noindent have been introduced in equation (2.1), while $\widehat{N}\left( 
\widehat{\mathbf{x}},\widehat{\mathbf{p}}^{\left( k\right) };\theta \right) $
is the smoothed microscopic phase space density

$$
\widehat{N}\left( \widehat{\mathbf{x}},\widehat{\mathbf{p}}^{\left( k\right)
};\theta \right) =\int d^3{\vec {\rho}}{\mathcal{G}}\left( \left. % 
\widehat{\mathbf{x}}%
\right| {\vec {\rho}}\right) N\left( {\vec {\rho }},\widehat{\mathbf{p%
}}^{\left( k\right) };\theta \right)  \eqno (2.4f)
$$

\noindent with a smoothing function ${\mathcal{G}}\left( \left. \mathbf{x}%
\right| \vec {\rho }\right) $.

The next step consists in averaging the Klimontovich equation (2.1) over the
relevant Gibbs ensemble with using the definition of one-particle
distribution function \cite{KlimSM}

$$
\left\langle N\left( \widehat{\mathbf{x}},\widehat{\mathbf{p}}^{\left(
k\right) };\theta \right) \right\rangle =nf\left( \widehat{\mathbf{x}},%
\widehat{\mathbf{p}}^{\left( k\right) };\theta \right) \qquad ,\qquad \left(
n=N_p/V\right)  \eqno (2.5a)
$$

$$
\int d^3\widehat{\mathbf{x}}d^3\widehat{\mathbf{p}}^{\left( k\right)
}f\left( \widehat{\mathbf{x}},\widehat{\mathbf{p}}^{\left( k\right) };\theta
\right) =V,  \eqno (2.5b)
$$

\noindent where $N_p$ is the total number of particles in the beam and $V$
is the volume occupied by the beam. By taking into account the
representation of the microscopic phase space density and the microscopic
force in terms of mean and fluctuating part

$$
N=nf+\delta N\quad ;\quad \widehat{\mathbf{F}}^{\left( M\right)
}=\left\langle \widehat{\mathbf{F}}\right\rangle +\delta \widehat{\mathbf{F}}%
\quad ;\quad \left\langle N\widehat{\mathbf{F}}^{\left( M\right)
}\right\rangle =nf\left\langle \widehat{\mathbf{F}}\right\rangle
+\left\langle \delta N\delta \widehat{\mathbf{F}}\right\rangle  \eqno (2.6)
$$

\noindent we obtain the generalized kinetic equation

$$
\frac{\partial f}{\partial \theta }+R\left( \widehat{\mathbf{v}}\cdot 
\widehat{\mathbf{\nabla }}_x\right) f+R\left[ \widehat{\mathbf{\nabla }}%
_p\cdot \left( \widehat{\mathbf{F}}_0+\left\langle \widehat{\mathbf{F}}%
\right\rangle \right) \right] f={\mathcal{J}}_{col}\left( \widehat{\mathbf{x}},%
\widehat{\mathbf{p}}^{\left( k\right) };\theta \right) +\widetilde{\mathcal{J%
}}\left( \widehat{\mathbf{x}},\widehat{\mathbf{p}}^{\left( k\right) };\theta
\right) ,  \eqno (2.7)
$$

\noindent where

$$
{\mathcal{J}}_{col}\left( \widehat{\mathbf{x}},\widehat{\mathbf{p}}^{\left(
k\right) };\theta \right) =-\frac Rn\widehat{\mathbf{\nabla }}_p\cdot
\left\langle \delta \widehat{\mathbf{F}}\delta N\right\rangle \qquad ;\qquad 
\widetilde{\mathcal{J}}\left( \widehat{\mathbf{x}},\widehat{\mathbf{p}}%
^{\left( k\right) };\theta \right) =\frac 1{\theta _{ph}}\left( \widehat{f}%
-f\right) .  \eqno (2.8)
$$

\noindent are the collision integrals. It was previously shown \cite{TzArc}, 
\cite{KlimS} that the additional collision integral $\widetilde{\mathcal{J}}%
\left( \widehat{\mathbf{x}},\widehat{\mathbf{p}}^{\left( k\right) };\theta
\right) $ can be cast into a Fokker-Planck ``collision term'', where the 
Fokker-Planck operator acts in coordinate space only. The equation for the 
fluctuating part $\delta N$ reads as

\[
\left[ \frac \partial {\partial \theta }+R\left( \widehat{\mathbf{v}}\cdot 
\widehat{\mathbf{\nabla }}_x\right) +R\widehat{\mathbf{\nabla }}_p\cdot
\left( \widehat{\mathbf{F}}_0+\left\langle \widehat{\mathbf{F}}\right\rangle
\right) \right] \delta N=-nR\widehat{\mathbf{\nabla }}_p\cdot \left( f\delta 
\widehat{\mathbf{F}}\right) + 
\]

$$
+R\widehat{\mathbf{\nabla }}_p\cdot \left[ \left\langle \delta \widehat{%
\mathbf{F}}\delta N\right\rangle -\delta \widehat{\mathbf{F}}\delta N\right]
+\frac 1{\theta _{ph}}\left( \widehat{\delta N}-\delta N\right) . 
\eqno (2.9)
$$

\noindent Averaging the Maxwell-Lorentz equations we get

$$
{\mathbf{\nabla }}_r\times \left\langle {\mathbf{B}}\right\rangle =\frac 1{c^2}%
\frac{\partial \left\langle {\mathbf{E}}\right\rangle }{\partial t}+\mu _0qn%
{\mathbf{j}}\left( {\mathbf{r}};t\right) \quad ;\quad {\mathbf{\nabla }}_r\times
\left\langle {\mathbf{E}}\right\rangle =-\frac{\partial \left\langle {\mathbf{B}}%
\right\rangle }{\partial t},  \eqno (2.10a)
$$

$$
{{\mathbf{\nabla }}_r\cdot \left\langle \mathbf{B}\right\rangle} =0\qquad
;\qquad {{\mathbf{\nabla }}_r\cdot \left\langle \mathbf{E}\right\rangle} =\frac{%
qn}{\varepsilon _0}\rho \left( {\mathbf{r}};t\right) ,  \eqno (2.10b)
$$

\noindent where

$$
\rho \left( {\mathbf{r}};t\right) =\int d^3{\mathbf{p}}^{\left( k\right)
}f\left( {\mathbf{r,p}}^{\left( k\right) };t\right) \qquad ;\qquad 
{\mathbf{j}}\left( {\mathbf{r}};t\right) =\int d^3{\mathbf{p}}^{\left( k\right) }%
{\mathbf{v}}f\left( {\mathbf{r,p}}^{\left( k\right) };t\right) . 
\eqno (2.11)
$$

\noindent The equations for the fluctuating fields are similar to (2.10) and
read as

$$
{\mathbf{\nabla }}_r\times \delta {\mathbf{B}}=\frac 1{c^2}\frac{\partial \delta 
{\mathbf{E}}}{\partial t}+\mu _0q\delta {\mathbf{j}}\left( {\mathbf{r}};t\right)
\quad ;\quad {\mathbf{\nabla }}_r\times \delta {\mathbf{E}}=-\frac{\partial
\delta {\mathbf{B}}}{\partial t},  \eqno (2.12a)
$$

$$
{\mathbf{\nabla }}_r\cdot \delta {\mathbf{B}}=0\qquad ;\qquad {\mathbf{\nabla }%
}_r\cdot \delta {\mathbf{E}}=\frac q{\varepsilon _0}\delta \rho \left( {\mathbf{%
r}};t\right) ,  \eqno (2.12b)
$$

\noindent where

$$
\delta \rho \left( {\mathbf{r}};t\right) =\int d^3{\mathbf{p}}^{\left( k\right)
}\delta N\left( {\mathbf{r,p}}^{\left( k\right) };t\right) \qquad
;\qquad \delta {\mathbf{j}}\left( {\mathbf{r}};t\right) =\int d^3{\mathbf{p}}%
^{\left( k\right) }{\mathbf{v}}\delta N\left( {\mathbf{r,p}}^{\left( k\right) }%
;t\right) .  \eqno (2.13)
$$

\noindent Taking divergence of the first of equations (2.12) and utilizing
the last one, it can be easily seen that the continuity equation for
fluctuating quantities holds

$$
\frac \partial {\partial t}\delta \rho \left( {\mathbf{r}};t\right) +{\mathbf{%
\nabla }}_r\cdot \delta {\mathbf{j}}\left( {\mathbf{r}};t\right) =0\qquad \quad
\left( \varepsilon _0\mu _0=1/c^2\right) .  \eqno (2.14)
$$

It should be pointed out that the microscopic electromagnetic fields depend
on the coordinates ${\mathbf{x}}=\left( x,z,\sigma \right) $ through the
microscopic phase space density $N$ written in these coordinates. The rest
of this section is dedicated to the derivation of some useful relations,
needed for the subsequent exposition. Consider the simple change of variables

\[
d^3{\mathbf{r}}d^3{\mathbf{p}}^{\left( k\right) }=\left( 1+{\mathbf{x\cdot K}}%
\right) ^2dxdzdsdp_x^{\left( k\right) }dp_z^{\left( k\right) }dp_s^{\left(
k\right) }= 
\]

\[
=\left( 1+{\mathbf{x\cdot K}}\right) ^2\left| \det {\mathcal{J}}_1\right|
dxdzd\sigma d\widetilde{p}_x^{\left( k\right) }d\widetilde{p}_z^{\left(
k\right) }dh^{\left( k\right) }. 
\]

\noindent Noting that

\[
x=\widetilde{x}\quad ;\quad z=\widetilde{z}\quad ;\quad s=\sigma +v_ot, 
\]

\[
p_x^{\left( k\right) }=p_o\widetilde{p}_x^{\left( k\right) }\quad ;\quad
p_z^{\left( k\right) }=p_o\widetilde{p}_z^{\left( k\right) }\quad ;\quad
p_s^{\left( k\right) }=\frac{p_o{\mathcal{S}}}{1+{\mathbf{x\cdot K}}}, 
\]

\[
{\mathcal{S}}=\sqrt{\beta _o^2h^{\left( k\right) 2}-\frac 1{\beta _o^2\gamma
_o^2}-\widetilde{p}_x^{\left( k\right) 2}-\widetilde{p}_z^{\left( k\right) 2}%
} 
\]

\noindent we easily find

\[
\left| \det {\mathcal{J}}_1\right| =p_o^3\frac{\beta _o^2h^{\left( k\right) }}{%
{\mathcal{S}}\left( 1+{\mathbf{x\cdot K}}\right) }. 
\]

\noindent Hence

\[
d^3{\mathbf{r}}d^3{\mathbf{p}}^{\left( k\right) }=p_o^3\left( 1+%
{\mathbf{x\cdot K}}%
\right) \frac{\beta _o^2h^{\left( k\right) }}{\mathcal{S}}dxdzd\sigma d%
\widetilde{p}_x^{\left( k\right) }d\widetilde{p}_z^{\left( k\right)
}dh^{\left( k\right) }. 
\]

\noindent Continuing further we use the relations

\[
u=\widehat{u}+\widehat{\eta }{\mathcal{D}}_u\quad ;\quad \widetilde{p}%
_u^{\left( k\right) }=\widehat{p}_u^{\left( k\right) }+\widehat{\eta }\frac{d%
{\mathcal{D}}_u}{ds}\quad ;\quad \sigma =\widehat{\sigma }+\sum_{u=\left(
x,z\right) }\left( \widehat{p}_u{\mathcal{D}}_u-\widehat{u}\frac{d{\mathcal{D}}_u%
}{ds}\right) , 
\]

\[
h^{\left( k\right) }=\widehat{\eta }^{\left( k\right) }+\frac 1{\beta _o^2} 
\]

\noindent and finally get

\[
d^3{\mathbf{r}}d^3{\mathbf{p}}^{\left( k\right) }=p_o^3\left( 1+%
{\mathbf{x\cdot K}}%
\right) \frac{1+\beta _o^2\widehat{\eta }^{\left( k\right) }}{\mathcal{S}}d%
\widehat{x}d\widehat{z}d\widehat{\sigma }d\widehat{p}_x^{\left( k\right) }d%
\widehat{p}_z^{\left( k\right) }d\widehat{\eta }^{\left( k\right) }. 
\]

\noindent As far as

\[
{\mathcal{S}}\approx \sqrt{1+2\widehat{\eta }^{\left( k\right) }+\beta _o^2%
\widehat{\eta }^{\left( k\right) 2}}\approx 1+\widehat{\eta }^{\left(
k\right) } 
\]

\noindent for $\beta _o\approx 1$ we obtain

$$
d^3{\mathbf{r}}d^3{\mathbf{p}}^{\left( k\right) }=p_o^3\left( 1+%
{\mathbf{x\cdot K}}%
\right) d^3\widehat{\mathbf{x}}d^3\widehat{\mathbf{p}}^{\left( k\right) }. 
\eqno (2.15)
$$

\noindent Thus, integration in the expressions for the charge and current
density

$$
\delta \rho \left( {\mathbf{r}};t\right) =\int d^3\widehat{\mathbf{p}}^{\left(
k\right) }\delta N\left( {\mathbf{x}},\widehat{\mathbf{p}}^{\left( k\right) }%
;\theta \right) \qquad ;\qquad \delta {\mathbf{j}}\left( {\mathbf{r}};%
t\right) =\int d^3\widehat{\mathbf{p}}^{\left( k\right) }{\mathbf{v}}\delta
N\left( {\mathbf{x}},\widehat{\mathbf{p}}^{\left( k\right) };\theta
\right) .  \eqno (2.16)
$$

\noindent goes approximately over the new kinetic momenta $\widehat{\mathbf{p%
}}^{\left( k\right) }$.

\section{Spectral Densities of Fluctuations.}

In order to determine the collision integral (2.8) we have to solve equation
(2.9) governing the evolution of fluctuations $\delta N$. Under the
assumption that fluctuations are small the second term on the right hand
side of equation (2.9) can be neglected

\[
\left[ \frac \partial {\partial \theta }+R\left( \widehat{\mathbf{v}}\cdot 
\widehat{\mathbf{\nabla }}_x\right) +R\widehat{\mathbf{\nabla }}_p\cdot
\left( \widehat{\mathbf{F}}_0+\left\langle \widehat{\mathbf{F}}\right\rangle
\right) \right] \delta N\left( \widehat{\mathbf{x}};\theta \right) = 
\]

$$
=-nR\widehat{\mathbf{\nabla }}_p\cdot \left( \delta \widehat{\mathbf{F}}%
\left( {\mathbf{x}};\theta \right) f\left( \widehat{\mathbf{x}};\theta \right)
\right) .  \eqno (3.1)
$$

\noindent The small source in the initial equation (2.9) has been dropped
off as non relevant for the dynamics of small-scale fluctuations. The term
containing the mean force in equation (3.1) can be neglected. This is
justified when calculating the small-scale fluctuations if

$$
\omega _p\gg \nu _{x,z,\sigma }\omega _o\qquad \quad \left( \omega _p^2=%
\frac{q^2n}{\varepsilon _0m_o}\quad ;\quad r_D^2=\frac{\varepsilon _0k_BT}{%
q^2n}\right) .  \eqno (3.2)
$$

\noindent Here $T$ is the temperature of the beam, $\omega _o$ is the
angular frequency of synchronous particle, $\nu _{x,z,\sigma }$ stands for
the betatron tunes in the two transverse planes as well as for the
synchrotron tune. Furthermore $\omega _p$ is the beam plasma frequency and $%
r_D$ - the Debye radius. It is worthwhile to note that the physical meaning
of Debye radius for particle beams is somewhat different from that commonly
used in plasma physics. In fact Debye radius is an equilibrium
characteristic of the beam, indicating the exponential decay of the
self-field, needed to self-maintain this equilibrium state.

The contribution of small-scale fluctuations can be better extracted if a
small source proportional to $\Delta $ is introduced into the left hand side
of (3.1)

$$
\left[ \frac \partial {\partial \theta }+R\left( \widetilde{\mathbf{v}}\cdot 
{\mathbf{\nabla }}_x\right) +\Delta \right] \delta N\left( {\mathbf{x}};\theta
\right) =-nR\widehat{\mathbf{\nabla }}_p\cdot \left( \delta \widehat{\mathbf{%
F}}\left( {\mathbf{x}};\theta \right) f\left( \widehat{\mathbf{x}};\theta
\right) \right) ,  \eqno (3.3)
$$

\[
\widetilde{\mathbf{v}}=\left( \widetilde{p}_x^{\left( k\right) },\widetilde{p%
}_z^{\left( k\right) },-{\mathcal{K}}\widehat{\eta }^{\left( k\right) }\right)
. 
\]

\noindent In going over from equation (3.1) to (3.3) the left hand side has
been represented in terms of the variables ${\mathbf{x}}=\left( x,z,\sigma
\right) $. The general solution of the above equation can be written as

$$
\delta N\left( {\mathbf{x}},\widehat{\mathbf{p}}^{\left( k\right) };\theta
\right) =\delta N^s\left( {\mathbf{x}},\widehat{\mathbf{p}}^{\left( k\right)
};\theta \right) +\delta N^{ind}\left( {\mathbf{x}},\widehat{\mathbf{p}}%
^{\left( k\right) };\theta \right) ,  \eqno (3.4)
$$

\noindent where $\delta N^{ind}$ is a generic solution of (3.3), while $%
\delta N^s$ accounts for the discrete structure of the beam as a collection
of particles. The latter can be determined from \cite{KlimSM}

$$
\left[ \frac \partial {\partial \theta }+R\left( \widetilde{\mathbf{v}}\cdot 
{\mathbf{\nabla }}_x\right) +\Delta \right] \left\langle \delta N^s\left( 
{\mathbf{X}};\theta \right) \delta N^s\left( {\mathbf{X}}_1;\theta
_1\right) \right\rangle =0\quad ;\quad \left( \mathbf{X}={\mathbf{x}},\widehat{%
\mathbf{p}}^{\left( k\right) }\right)  \eqno (3.5)
$$

\noindent with the initial condition

$$
\left\langle \delta N^s\left( {\mathbf{X}};\theta \right) \delta N^s\left( 
{\mathbf{X}}_1\mathbf{;}\theta \right) \right\rangle =n\delta \left( {\mathbf{%
x-x}}_1\right) \delta \left( \widehat{\mathbf{p}}^{\left( k\right) }-\widehat{%
\mathbf{p}}_1^{\left( k\right) }\right) f\left( {\mathbf{x}},\widehat{\mathbf{p%
}}^{\left( k\right) };\theta \right) .  \eqno (3.6)
$$

\noindent When small-scale fluctuations are computed $f\left( {\mathbf{x}},%
\widehat{\mathbf{p}}^{\left( k\right) };\theta \right) $ can be considered a
smooth enough function (not varying considerably) and $\left\langle \delta
N^s\left( {\mathbf{X}};\theta \right) \delta N^s\left( {\mathbf{X}}_1;%
\theta _1\right) \right\rangle $ depends on $\theta -\theta _1$ and ${\mathbf{%
x-x}}_1$ only. Introducing the Fourier transform:

\[
\left\langle \delta N\left( {\mathbf{X}};\theta \right) \delta N\left( {\mathbf{%
X}}_1\;\theta _1\right) \right\rangle =\left\langle \delta N\delta
N\right\rangle \left( \theta -\theta _1,{\mathbf{x-x}}_1,\widehat{\mathbf{p}}%
^{\left( k\right) },\widehat{\mathbf{p}}_1^{\left( k\right) }\right) = 
\]

$$
=\frac 1{\left( 2\pi \right) ^3}\int d^3{\mathbf{k}}\left( \widetilde{\delta
N\delta N}\right) \left( \theta -\theta _1,{\mathbf{k}},\widehat{\mathbf{p}}%
^{\left( k\right) },\widehat{\mathbf{p}}_1^{\left( k\right) }\right) \exp
\left[ i{\mathbf{k\cdot }}\left( {\mathbf{x-x}}_1\right) \right]  \eqno (3.7)
$$

\noindent we cast equation (3.5) into the form

$$
\left( \frac \partial {\partial \theta }+iR{\mathbf{k\cdot }}\widetilde{%
\mathbf{v}}+\Delta \right) \left( \widetilde{\delta N\delta N}\right)
^s\left( \tau ,{\mathbf{k}},\widehat{\mathbf{p}}^{\left( k\right) },\widehat{%
\mathbf{p}}_1^{\left( k\right) }\right) =0\qquad \left( \tau =\theta -\theta
_1\right) ,  \eqno (3.5a)
$$

$$
\left( \widetilde{\delta N\delta N}\right) ^s\left( \left. \tau ,{\mathbf{k}},%
\widehat{\mathbf{p}}^{\left( k\right) },\widehat{\mathbf{p}}_1^{\left(
k\right) }\right) \right| _{\tau =0}=n\delta \left( \widehat{\mathbf{p}}%
^{\left( k\right) }-\widehat{\mathbf{p}}_1^{\left( k\right) }\right) f\left( 
{\mathbf{x}},\widehat{\mathbf{p}}^{\left( k\right) };\theta \right) . 
\eqno (3.6a)
$$

\noindent Further we introduce the one-sided Fourier transform in the time
domain

$$
\left( \widetilde{\delta N\delta N}\right) ^{\dagger }\left( \omega ,{\mathbf{%
k}},\widehat{\mathbf{p}}^{\left( k\right) },\widehat{\mathbf{p}}_1^{\left(
k\right) }\right) =\int\limits_0^\infty d\tau \left( \widetilde{\delta
N\delta N}\right) \left( \tau ,{\mathbf{k}},\widehat{\mathbf{p}}^{\left(
k\right) },\widehat{\mathbf{p}}_1^{\left( k\right) }\right) \exp \left(
i\omega \tau \right) .  \eqno (3.8)
$$

\noindent Multiplication of equation (3.5a) by $e^{i\omega \tau }$ and
subsequent integration on $\tau $ yields:

\[
\left( -i\omega +iR{\mathbf{k\cdot }}\widetilde{\mathbf{v}}+\Delta \right)
\left( \widetilde{\delta N\delta N}\right) ^{\dagger }\left( \omega ,{\mathbf{%
k}},\widehat{\mathbf{p}}^{\left( k\right) },\widehat{\mathbf{p}}_1^{\left(
k\right) }\right) =\left( \widetilde{\delta N\delta N}\right) ^s\left( 0,%
{\mathbf{k}},\widehat{\mathbf{p}}^{\left( k\right) },\widehat{\mathbf{p}}%
_1^{\left( k\right) }\right) , 
\]

\noindent or

$$
\left( \widetilde{\delta N\delta N}\right) ^{\dagger }\left( \omega ,{\mathbf{%
k}},\widehat{\mathbf{p}}^{\left( k\right) },\widehat{\mathbf{p}}_1^{\left(
k\right) }\right) =\frac{inf\left( {\mathbf{x}},\widehat{\mathbf{p}}^{\left(
k\right) };\theta \right) }{\omega -R{\mathbf{k\cdot }}\widetilde{\mathbf{v}}%
+i\Delta }\delta \left( \widehat{\mathbf{p}}^{\left( k\right) }-\widehat{%
\mathbf{p}}_1^{\left( k\right) }\right) .  \eqno (3.9)
$$

\noindent Using the equation

\[
\left( \widetilde{\delta N\delta N}\right) \left( \omega ,{\mathbf{k}},%
\widehat{\mathbf{p}}^{\left( k\right) },\widehat{\mathbf{p}}_1^{\left(
k\right) }\right) =\left( \widetilde{\delta N\delta N}\right) ^{\dagger
}\left( \omega ,{\mathbf{k}},\widehat{\mathbf{p}}^{\left( k\right) },\widehat{%
\mathbf{p}}_1^{\left( k\right) }\right) + 
\]

\[
+\left[ \left( \widetilde{\delta N\delta N}\right) ^{\dagger }\left( \omega ,%
{\mathbf{k}},\widehat{\mathbf{p}}^{\left( k\right) },\widehat{\mathbf{p}}%
_1^{\left( k\right) }\right) \right] _{\widehat{\mathbf{p}}\leftrightarrow 
\widehat{\mathbf{p}}_1}^{*} 
\]

\noindent relating the one-sided and two-sided Fourier transform we get

\[
\left( \widetilde{\delta N\delta N}\right) ^s\left( \omega ,{\mathbf{k}},%
\widehat{\mathbf{p}}^{\left( k\right) },\widehat{\mathbf{p}}_1^{\left(
k\right) }\right) =\frac{2\Delta }{\left( \omega -R{\mathbf{k\cdot }}%
\widetilde{\mathbf{v}}\right) ^2+\Delta ^2}nf\left( {\mathbf{x}},\widehat{%
\mathbf{p}}^{\left( k\right) };\theta \right) \delta \left( \widehat{\mathbf{%
p}}^{\left( k\right) }-\widehat{\mathbf{p}}_1^{\left( k\right) }\right) . 
\]

\noindent The definition of Dirac's $\delta $-function

\[
\lim\limits_{\Delta \rightarrow 0}\frac \Delta {\left( \omega -R{\mathbf{%
k\cdot }}\widetilde{\mathbf{v}}\right) ^2+\Delta ^2}=\pi \delta \left( \omega
-R{\mathbf{k\cdot }}\widetilde{\mathbf{v}}\right) 
\]

\noindent gives finally

$$
\left( \widetilde{\delta N\delta N}\right) ^s\left( \omega ,{\mathbf{k}},%
\widehat{\mathbf{p}}^{\left( k\right) },\widehat{\mathbf{p}}_1^{\left(
k\right) }\right) =2\pi nf\left( {\mathbf{x}},\widehat{\mathbf{p}}^{\left(
k\right) };\theta \right) \delta \left( \widehat{\mathbf{p}}^{\left(
k\right) }-\widehat{\mathbf{p}}_1^{\left( k\right) }\right) \delta \left(
\omega -R{\mathbf{k\cdot }}\widetilde{\mathbf{v}}\right) .  \eqno (3.10)
$$

\noindent To obtain an arbitrary solution of equation (3.3) we perform the
Fourier transform

\[
\delta N\left( {\mathbf{x}},\widehat{\mathbf{p}}^{\left( k\right) };\theta
\right) =\frac 1{\left( 2\pi \right) ^4}\int d\omega d^3{\mathbf{k}}\delta 
\widetilde{N}\left( \omega ,{\mathbf{k}},\widehat{\mathbf{p}}^{\left( k\right)
}\right) e^{i\left( {\mathbf{k\cdot x}}-\omega \theta \right) } 
\]

\[
\delta \widetilde{N}\left( \omega ,{\mathbf{k}},\widehat{\mathbf{p}}^{\left(
k\right) }\right) =\int d\theta d^3{\mathbf{x}}\delta N\left( {\mathbf{x}},%
\widehat{\mathbf{p}}^{\left( k\right) };\theta \right) e^{i\left( \omega
\theta -{\mathbf{k\cdot x}}\right) } 
\]

\noindent and find

$$
\delta \widetilde{N}\left( \omega ,{\mathbf{k}},\widehat{\mathbf{p}}^{\left(
k\right) }\right) =\delta \widetilde{N}^s\left( \omega ,{\mathbf{k}},\widehat{%
\mathbf{p}}^{\left( k\right) }\right) -\frac{inR}{\omega -R{\mathbf{k\cdot }}%
\widetilde{\mathbf{v}}+i\Delta }\widehat{\mathbf{\nabla }}_p\cdot \left[ 
\widetilde{\delta \widehat{\mathbf{F}}}\left( \omega, {\mathbf{k}}\right)
f\left( \widehat{\mathbf{x}};\theta \right) \right] .  \eqno (3.11)
$$

What remains now is to compute the spectral density of fluctuating force $%
\widetilde{\delta \widehat{\mathbf{F}}}$. In doing so we consider an
arbitrary function $F\left( {\mathbf{x}};\theta \right) $. Let the same
function, written in the variables ${\mathbf{r}}=\left( x,z,s=R\theta \right) $
and $t$ be $F_r\left( {\mathbf{r}};t\right) $. Further we have

\[
F_r\left( {\mathbf{r}};t\right) =\frac 1{\left( 2\pi \right) ^4}\int d\nu d^3%
{\mathbf{m}}\widetilde{F}_r\left( \nu ;{\mathbf{m}}\right) e^{i\left( {\mathbf{%
m\cdot r}}-\nu t\right) }= 
\]

\[
=\frac 1{\left( 2\pi \right) ^4}\int d\nu d^3{\mathbf{m}}\widetilde{F}_r\left(
\nu ;{\mathbf{m}}\right) \exp \left\{ i\left[ m_xx+m_zz+m_sR\theta -\frac{\nu
\left( R\theta -\sigma \right) }{v_o}\right] \right\} = 
\]

\[
=\frac{\omega _o}{\left( 2\pi \right) ^4}\int d\omega d^3{\mathbf{k}}%
\widetilde{F}_r\left( v_ok_\sigma ;k_x,k_z,k_\sigma -\frac \omega R\right)
e^{i\left( {\mathbf{k\cdot x}}-\omega \theta \right) }, 
\]

\noindent where the following change of variables

$$
{\mathbf{m}}=\left( k_x,k_z,k_\sigma -\frac \omega R\right) \qquad ;\qquad \nu
=v_ok_\sigma  \eqno (3.12)
$$

\noindent has been introduced. Therefore the relation we are looking for
reads as

$$
\widetilde{F}\left( \omega ;{\mathbf{k}}\right) =\omega _o\widetilde{F}%
_r\left( v_ok_\sigma ;k_x,k_z,k_\sigma -\frac \omega R\right) .  \eqno (3.13)
$$

Fourier analysing equations (2.12) we find

$$
i{\mathbf{m}}\times \delta \widetilde{\mathbf{B}}_r=-\frac{i\nu }{c^2}\delta 
\widetilde{\mathbf{E}}_r+\mu _0q\delta \widetilde{\mathbf{j}}_r\qquad
;\qquad \delta \widetilde{\mathbf{B}}_r=\frac 1\nu {\mathbf{m}}\times \delta 
\widetilde{\mathbf{E}}_r,  \eqno (3.14a)
$$

$$
{\mathbf{m\cdot }}\delta \widetilde{\mathbf{B}}_r=0\qquad ;\qquad i{\mathbf{%
m\cdot }}\delta \widetilde{\mathbf{E}}_r=\frac q{\varepsilon _0}\delta 
\widetilde{\rho }_r.  \eqno (3.14b)
$$

\noindent Let us represent the electromagnetic fields as a sum of
longitudinal and transversal components

$$
\delta \widetilde{\mathbf{E}}_r=\delta \widetilde{\mathbf{E}}_r^{\parallel
}+\delta \widetilde{\mathbf{E}}_r^{\perp }\qquad \left( {\mathbf{m}}\times %
\delta \widetilde{\mathbf{E}}_r^{\parallel }=0\quad ;\quad {\mathbf{m\cdot }}%
\delta \widetilde{\mathbf{E}}_r^{\perp }=0\right)  \eqno (3.15)
$$

\noindent and further simplify the problem by considering

$$
\delta \widetilde{\mathbf{j}}_r=v_o{\mathbf{e}}_s\delta \widetilde{\rho }_r. 
\eqno (3.16)
$$

\noindent From the continuity equation (2.14) we get

$$
\delta \widetilde{\rho }_r=\frac 1\nu {\mathbf{m\cdot }}\delta \widetilde{%
\mathbf{j}}_r  \eqno (3.17)
$$

\noindent and using (3.13) and (3.16) we conclude that $\mathbf{m=k}$. Thus
we obtain

$$
\delta \widetilde{\mathbf{E}}^{\parallel }\left( \omega ,{\mathbf{k}}\right) =-%
\frac{iq\mathbf{k}}{\varepsilon _0k^2}\delta \widetilde{\rho }\left( \omega ,%
{\mathbf{k}}\right) ,  \eqno (3.18a)
$$

$$
\delta \widetilde{\mathbf{E}}^{\perp }\left( \omega ,{\mathbf{k}}\right) =%
\frac{iq\beta _o^2k_\sigma }{\varepsilon _0k^2\left( k^2-\beta _o^2k_\sigma
^2\right) }\left[ {\mathbf{k}}\times \left( {\mathbf{e}}_s\times {\mathbf{k}}%
\right) \right] \delta \widetilde{\rho }\left( \omega ,{\mathbf{k}}\right) , 
\eqno (3.18b)
$$

$$
\delta \widetilde{\mathbf{B}}\left( \omega ,{\mathbf{k}}\right) =\frac
1{v_ok_\sigma }{\mathbf{k}}\times \delta \widetilde{\mathbf{E}}^{\perp }\left(
\omega ,{\mathbf{k}}\right) =\frac{iq\beta _o^2}{\varepsilon _0v_o\left(
k^2-\beta _o^2k_\sigma ^2\right) }\left( {\mathbf{k\times e}}_s\right) \delta 
\widetilde{\rho }\left( \omega ,{\mathbf{k}}\right) .  \eqno (3.18c)
$$

\noindent Retaining leading terms only, we write the fluctuating force $%
\widetilde{\delta \widehat{\mathbf{F}}}$ as

$$
\widetilde{\delta \widehat{\mathbf{F}}}\left( \omega ,{\mathbf{k}}\right)
=\frac q{\beta _o^2E_o}\left[ \delta \widetilde{\mathbf{E}}+v_o\left( 
{\mathbf{e}}_s\times \delta \widetilde{\mathbf{B}}\right) \right] =-\frac{iq^2%
\mathbf{k}}{\varepsilon _0\beta _o^2\gamma _o^2E_o\left( k^2-\beta
_o^2k_\sigma ^2\right) }\delta \widetilde{\rho }\left( \omega ,{\mathbf{k}}%
\right) .  \eqno (3.19)
$$

\noindent Integrating equation (3.11) on $\widehat{\mathbf{p}}^{\left(
k\right) }$ we obtain

\[
\delta \widetilde{\rho }\left( \omega ,\mathbf{k}\right) =\delta \widetilde{%
\rho }^s\left( \omega ,\mathbf{k}\right) -inR\int d^3\widehat{\mathbf{p}}%
^{\left( k\right) }\frac{\widehat{\mathbf{\nabla }}_pf\left( \widehat{%
\mathbf{x}};\theta \right) }{\omega -R{\mathbf{k\cdot }}\widetilde{\mathbf{v}}%
+i\Delta }\cdot \widetilde{\delta \widehat{\mathbf{F}}}\left( \omega ,%
\mathbf{k}\right) 
\]

\noindent and eliminating $\widetilde{\delta \widehat{\mathbf{F}}}\left(
\omega ,\mathbf{k}\right) $ with (3.19) in hand we get finally

$$
\widetilde{\epsilon }\left( \omega ,\mathbf{k}\right) \delta \widetilde{\rho 
}\left( \omega ,\mathbf{k}\right) =\delta \widetilde{\rho }^s\left( \omega ,%
\mathbf{k}\right) ,  \eqno (3.20)
$$

\noindent where

$$
\widetilde{\epsilon }\left( \omega ,\mathbf{k}\right) =1+\frac{q^2nR}{%
\varepsilon _0\beta _o^2\gamma _o^2E_o\left( k^2-\beta _o^2k_\sigma
^2\right) }\int d^3\widehat{\mathbf{p}}^{\left( k\right) }\frac{{\mathbf{%
k\cdot }}\widehat{\mathbf{\nabla }}_pf\left( \widehat{\mathbf{x}};\theta
\right) }{\omega -R{\mathbf{k\cdot }}\widetilde{\mathbf{v}}+i\Delta } 
\eqno (3.21)
$$

\noindent is the dielectric susceptibility of the beam. Thus for the
spectral density of the fluctuating force we have the following expression:

$$
\widetilde{\delta \widehat{\mathbf{F}}}\left( \omega ,\mathbf{k}\right) =-%
\frac{iq^2\mathbf{k}}{\varepsilon _0\widetilde{\epsilon }\left( \omega ,%
\mathbf{k}\right) \beta _o^2\gamma _o^2E_o\left( k^2-\beta _o^2k_\sigma
^2\right) }\delta \widetilde{\rho }^s\left( \omega ,\mathbf{k}\right) . 
\eqno (3.22)
$$

\section{Collision Integral in the Form of Balescu-Lenard.}

According to (2.8) the collision integral is given by

$$
{\mathcal{J}}_{col}\left( \widehat{\mathbf{x}},\widehat{\mathbf{p}}^{\left(
k\right) };\theta \right) =-\frac Rn\widehat{\mathbf{\nabla }}_p\cdot
\left\langle \delta \widehat{\mathbf{F}}\delta N\right\rangle \left( {\mathbf{%
x}},\widehat{\mathbf{p}}^{\left( k\right) },\theta ;{\mathbf{x}},\widehat{%
\mathbf{p}}^{\left( k\right) },\theta \right) .  \eqno (4.1)
$$

\noindent We shall express the right hand side of (4.1) in terms of the
spectral densities of fluctuations $\widetilde{\delta \widehat{\mathbf{F}}}$
and $\delta \widetilde{N}$. Let ${\mathcal{F}}\left( {\mathbf{x}};\theta \right) 
$ and ${\mathcal{G}}\left( {\mathbf{x}}_1;\theta _1\right) $ be two random
functions. The second moment in the variables ${\mathbf{x-x}}_1$, $\theta
-\theta _1$ can be written as

\[
\left\langle {\mathcal{FG}}\right\rangle \left( {\mathbf{x}},\theta ;{\mathbf{x}}%
_1,\theta _1\right) =\left\langle {\mathcal{FG}}\right\rangle \left( {\mathbf{x}}%
,\theta ;{\mathbf{x-x}}_1,\theta -\theta _1\right) = 
\]

$$
=\frac 1{\left( 2\pi \right) ^4}\int d\omega d^3{\mathbf{k}}\left( \widetilde{%
\mathcal{FG}}\right) \left( \omega ,{\mathbf{k;x,}}\theta \right) \exp \left\{
i\left[ {\mathbf{k\cdot }}\left( {\mathbf{x-x}}_1\right) -\omega \left( \theta
-\theta _1\right) \right] \right\} .  \eqno (4.2)
$$

\noindent As far as the second moment is a real function the spectral
density obeys

$$
\left( \widetilde{\mathcal{FG}}\right) \left( \omega ,{\mathbf{k;x,}}\theta
\right) =\left( \widetilde{\mathcal{FG}}\right) ^{*}\left( -\omega ,-{\mathbf{%
k;x,}}\theta \right) .  \eqno (4.3)
$$

\noindent Letting ${\mathbf{x=x}}_1$, $\theta =\theta _1$ in (4.2) with (4.3)
in hand we find

$$
\left\langle {\mathcal{FG}}\right\rangle \left( \mathbf{x,}\theta ;\mathbf{x}%
,\theta \right) =\frac 1{\left( 2\pi \right) ^4}\int d\omega d^3{\mathbf{k\ }}%
Re\left( \widetilde{\mathcal{FG}}\right) \left( \omega ,{\mathbf{k;x,}}\theta
\right) .  \eqno (4.4)
$$

\noindent Using (4.4) and taking into account only leading terms in $%
\widetilde{\delta \widehat{\mathbf{F}}}$ we rewrite (4.1) as

$$
{\mathcal{J}}_{col}\left( \widehat{\mathbf{x}},\widehat{\mathbf{p}}^{\left(
k\right) };\theta \right) =-\frac Rn\widehat{\mathbf{\nabla }}_p\cdot \int 
\frac{d\omega d^3{\mathbf{k}}}{\left( 2\pi \right) ^4}Re\left( \widetilde{%
\delta \widehat{\mathbf{F}}\delta N}\right) \left( \omega ,{\mathbf{k}};%
{\mathbf{x}},\widehat{\mathbf{p}}^{\left( k\right) },\theta \right) . 
\eqno (4.5)
$$

\noindent Utilizing the expressions (3.11) and (3.22) we obtain

\[
\left( \widetilde{\delta \widehat{\mathbf{F}}\delta N}\right) \left( \omega ,%
{\mathbf{k}};{\mathbf{x}},\widehat{\mathbf{p}}^{\left( k\right) },\theta \right)
=-\frac{inR{\mathbf{k\cdot }}\widehat{\mathbf{\nabla }}_pf\left( \widehat{%
\mathbf{x}};\theta \right) }{\omega -R{\mathbf{k\cdot }}\widetilde{\mathbf{v}}%
+i\Delta }{\mathbf{k}}\left( \widetilde{\delta \widehat{F}\delta \widehat{F}}%
\right) _{\omega ,{\mathbf{k}}}- 
\]

\[
-\frac{iq^2n\mathbf{k}}{\varepsilon _0\widetilde{\epsilon }\beta _o^2\gamma
_o^2E_o\left( k^2-\beta _o^2k_\sigma ^2\right) }f\left( \mathbf{x;}\theta
\right) 2\pi \delta \left( \omega -R\mathbf{k\cdot }\widetilde{\mathbf{v}}%
\right) , 
\]

\noindent where

$$
\left( \widetilde{\delta \widehat{F}\delta \widehat{F}}\right) _{\omega ,%
\mathbf{k}}=\frac{q^4n}{\varepsilon _0^2\left| \widetilde{\epsilon }\right|
^2\beta _o^4\gamma _o^4E_o^2\left( k^2-\beta _o^2k_\sigma ^2\right) ^2}\int
d^3\widehat{\mathbf{p}}^{\left( k\right) }f\left( \mathbf{x};\theta \right)
2\pi \delta \left( \omega -R\mathbf{k\cdot }\widetilde{\mathbf{v}}\right) . 
\eqno (4.6)
$$

\noindent In formula (4.5) representing the collision integral only the real
part of $\left( \widetilde{\delta \widehat{\mathbf{F}}\delta N}\right)
\left( \omega ,{\mathbf{k}};{\mathbf{x}},\widehat{\mathbf{p}}^{\left( k\right)
},\theta \right) $ enters. Therefore the expression to be substituted back
into (4.5) reads as

\[
Re\left( \widetilde{\delta \widehat{\mathbf{F}}\delta N}\right) \left(
\omega ,{\mathbf{k}};{\mathbf{x}},\widehat{\mathbf{p}}^{\left( k\right) },\theta
\right) =-\pi nR\delta \left( \omega -R{\mathbf{k\cdot }}\widetilde{\mathbf{v}}%
\right) {\mathbf{k}}\left( \widetilde{\delta \widehat{F}\delta \widehat{F}}%
\right) _{\omega ,{\mathbf{k}}}{\mathbf{k}}\cdot \widehat{\mathbf{\nabla }}%
_pf\left( \widehat{\mathbf{x}};\theta \right) - 
\]

$$
-\frac{2\pi q^2n{\mathbf{k}}}{\varepsilon _0\beta _o^2\gamma _o^2E_o\left(
k^2-\beta _o^2k_\sigma ^2\right) }\delta \left( \omega -R{\mathbf{k\cdot }}%
\widetilde{\mathbf{v}}\right) \frac{Im\widetilde{\epsilon }\left( \omega ,%
{\mathbf{k}}\right) }{\left| \widetilde{\epsilon }\left( \omega ,\mathbf{k}%
\right) \right| ^2}f\left( {\mathbf{x}};\theta \right) ,  \eqno (4.7)
$$

\noindent where

$$
Im\widetilde{\epsilon }\left( \omega ,{\mathbf{k}}\right) =-\frac{\pi q^2nR}{%
\varepsilon _0\beta _o^2\gamma _o^2E_o\left( k^2-\beta _o^2k_\sigma
^2\right) }\int d^3\widehat{\mathbf{p}}^{\left( k\right) }\delta \left(
\omega -R{\mathbf{k\cdot }}\widetilde{\mathbf{v}}\right) {\mathbf{k}}\cdot %
\widehat{\mathbf{\nabla }}_pf\left( \widehat{\mathbf{x}};\theta \right) . 
\eqno (4.8)
$$

\noindent Finally, the collision integral (4.5) can be written in the form
of Balescu-Lenard as

\[
{\mathcal{J}}_{col}^{\left( BL\right) }\left( \widehat{\mathbf{x}},\widehat{%
\mathbf{p}}^{\left( k\right) };\theta \right) =\frac{\pi q^4nR}{\varepsilon
_0^2\beta _o^4\gamma _o^4E_o^2}\widehat{\mathbf{\nabla }}_p\cdot \int \frac{%
d^3{\mathbf{k}}d^3\widehat{\mathbf{p}}_1^{\left( k\right) }}{\left( 2\pi
\right) ^3}\delta \left( {\mathbf{k}}\cdot \widetilde{\mathbf{v}}-{\mathbf{%
k}}\cdot \widetilde{\mathbf{v}}_1\right) * 
\]

\[
\ast \frac{\mathbf{kk}}{\left| \widetilde{\epsilon }\left( R\mathbf{k\cdot }%
\widetilde{\mathbf{v}},\mathbf{k}\right) \right| ^2\left( k^2-\beta
_o^2k_\sigma ^2\right) ^2}\cdot 
\]

$$
\cdot \left[ f\left( {\mathbf{x}},\widehat{\mathbf{p}}_1^{\left( k\right)
};\theta \right) \widehat{\mathbf{\nabla }}_pf\left( \widehat{\mathbf{x}}%
,\widehat{\mathbf{p}}^{\left( k\right) };\theta \right) -f\left( 
{\mathbf{x}},\widehat{\mathbf{p}}^{\left( k\right) };\theta \right) \widehat{%
\mathbf{\nabla }}_{p_1}f\left( \widehat{\mathbf{x}},\widehat{%
\mathbf{p}}_1^{\left( k\right) };\theta \right) \right] .  \eqno (4.9)
$$

\noindent The collision integral (4.5) can be put in an equivalent form of a
nonlinear Fokker-Planck operator

$$
{\mathcal{J}}_{col}^{\left( BL\right) }\left( \widehat{\mathbf{x}},\widehat{%
\mathbf{p}}^{\left( k\right) };\theta \right) =\widehat{\mathbf{\nabla }}%
_p\cdot \left[ \widehat{\mathcal{D}}^{\left( BL\right) }\cdot \widehat{%
\mathbf{\nabla }}_pf\left( \widehat{\mathbf{x}},\widehat{\mathbf{p}}%
^{\left( k\right) };\theta \right) \right] +\widehat{\mathbf{\nabla }}%
_p\cdot \left[ {\mathbf{A}}^{\left( BL\right) }f\left( {\mathbf{x}},\widehat{%
\mathbf{p}}^{\left( k\right) };\theta \right) \right] ,  \eqno (4.10)
$$

\noindent where the drift and diffusion coefficients

$$
\widehat{\mathcal{D}}^{\left( BL\right) }=\pi R^2\int \frac{d\omega d^3%
{\mathbf{k}}}{\left( 2\pi \right) ^4}\delta \left( \omega -R{\mathbf{k}}\cdot %
\widetilde{\mathbf{v}}\right) {\mathbf{k}}\left( \widetilde{\delta \widehat{F}%
\delta \widehat{F}}\right) _{\omega ,\mathbf{k}}{\mathbf{k}},  \eqno (4.11a)
$$

$$
{\mathbf{A}}^{\left( BL\right) } = \frac{q^2R}{\varepsilon _0\beta
_o^2\gamma _o^2E_o}\int \frac{d\omega d^3{\mathbf{k}}}{\left( 2\pi \right) ^4}%
\delta \left( \omega -R{\mathbf{k}}\cdot \widetilde{\mathbf{v}}\right) \frac{%
{\mathbf{k}}}{k^2-\beta _o^2k_\sigma ^2}\frac{Im\widetilde{\epsilon }\left(
\omega ,{\mathbf{k}}\right) }{\left| \widetilde{\epsilon }\left( \omega ,%
{\mathbf{k}}\right) \right| ^2}  \eqno (4.11b)
$$

\noindent depend on the distribution function itself.

\section{Collision Integral in the Form of Landau.}

The dielectric function (3.21) depends on the distribution function and
consequently the corresponding kinetic equation with the collision integral
in the form of Balescu-Lenard is extremely complicated to solve. Thus one
should seek reasonable ways for further simplifications. First of all we
shall determine the equilibrium state described by $f_0\left( \widehat{%
\mathbf{x}},\widehat{\mathbf{p}}^{\left( k\right) };\theta \right) $
and satisfying

$$
\left[ \frac \partial {\partial \theta }+R\left( \widehat{\mathbf{v}}\cdot 
\widehat{\mathbf{\nabla }}_x\right) +R\left( \widehat{\mathbf{F}}_0\cdot 
\widehat{\mathbf{\nabla }}_p\right) \right] f_0\left( \widehat{\mathbf{x}}%
,\widehat{\mathbf{p}}^{\left( k\right) };\theta \right) =0. 
\eqno (5.1)
$$

\noindent It can be easily checked that equation (5.1) has a solution of the
form

$$
f_0\left( \widehat{\mathbf{x}},\widehat{\mathbf{p}}^{\left(
k\right) };\theta \right) = f_0\left( \frac{2J_x}{2\epsilon _x},\frac{2J_z}{%
2\epsilon _z},\frac{2J_\sigma }{2\epsilon _\sigma }\right) ,  \eqno (5.2)
$$

\noindent where

$$
2J_u=\frac 1{\beta _u}\left[ \widehat{u}^2+\left( \beta _u\widehat{p}%
_u^{\left( k\right) }+\alpha _u\widehat{u}\right) ^2\right] \qquad ;\qquad
\left( u=x,z\right) ,  \eqno (5.3a)
$$

$$
2J_\sigma =\frac{\widehat{\eta }^{\left( k\right) 2}}\lambda +\lambda \left( 
\widehat{\sigma }-\sigma _s+\frac R\kappa \tan \Phi _s\right) ^2.  \eqno (5.3b)
$$

\noindent In the above expressions $\alpha $, $\beta $ and $\gamma $ are the
well-known Twiss parameters

\[
\frac{d\alpha _u}{d\theta }=\frac{G_u}R\beta _u-R\gamma _u\quad ;\quad \frac{%
d\beta _u}{d\theta }=-2R\alpha _u\quad ;\quad \frac{d\gamma _u}{d\theta }=%
\frac{2G_u}R\alpha _u, 
\]

\[
\beta _u\gamma _u-\alpha _u^2=1, 
\]

\noindent while

\[
\lambda ^2=\frac 1{2\pi R^2}\frac{\Delta E_0}{\beta _o^2E_o}\frac \kappa {%
\mathcal{K}}\cos \Phi _s\quad ;\quad \left( \beta _\sigma =\lambda
^{-1}\right) \quad ;\quad \nu _s^2=R^2{\mathcal{K}}^2\lambda ^2, 
\]

\noindent $\kappa $ - being the harmonic acceleration number, $\Phi _s$ - the
phase of synchronous particle, $\nu _s$ is the synchrotron tune and $\beta
_\sigma $ can be interpreted as the ''synchrotron beta-function''. The
quantities $\epsilon _x$, $\epsilon _z$ and $\epsilon _\sigma $ are related
to the transverse and longitudinal beam size and are referred to as
equilibrium beam emittances. To describe a local equilibrium state (see next
section) one can formally choose the equilibrium beam emittances $\epsilon
_x $, $\epsilon _z$ and $\epsilon _\sigma $ proportional to the
beta-functions by a universal scaling factor $\epsilon /R$ characterizing
the equilibrium state. Let us recall that at local equilibrium all the
parameters of the distribution are allowed to depend on coordinates and time 
\cite{KlimSM}, which is consistent with the specific choice above. Further,
by specifying the generic function (5.2) for slowly varying beam envelopes
we find

$$
f_0\left( \widehat{\mathbf{x}},\widehat{\mathbf{p}}^{\left(
k\right) };\theta \right) =V\left( \frac R{2\pi \epsilon }\right) ^{3/2}\exp
\left[ -\frac{2R\left( J_x\beta _x^{-1}+J_z\beta _z^{-1}+J_\sigma \beta
_\sigma ^{-1}\right) }{2\epsilon }\right] .  \eqno (5.4)
$$

\noindent The equilibrium beam emittances $\epsilon _x$, $\epsilon _z$ and $%
\epsilon _\sigma $ are related to the temperature of the beam through the
expression

$$
\epsilon _{x,z,\sigma }=\frac \epsilon R\beta _{x,z,\sigma }\qquad ;\qquad
\left( \frac \epsilon R=\frac{k_BT}{\beta _o^2E_o}\right)  \eqno (5.5)
$$

In order to obtain the collision integral in the form of Landau we consider $%
\widetilde{\epsilon }\left( R{\mathbf{k}}\cdot \widetilde{\mathbf{v}},{\mathbf{k%
}}\right) =1$ in equation (4.9) and simultaneously take into account the
effect of polarization by altering the domain of integration on $k$ for
small $k$. As far as the large values of $k$ are concerned the upper limit
of integration can be obtained from the condition that perturbation
expansion holds. To proceed further it is convenient to change variables in
the Balescu-Lenard kinetic equation according to

\[
\widehat{\eta }^{\left( k\right) }\longrightarrow -sign\left( \mathcal{K}%
\right) \frac{\widehat{\eta }^{\left( k\right) }}{\sqrt{\left| \mathcal{K}%
\right| }}\qquad ;\qquad k_\sigma \longrightarrow \frac{k_\sigma }{\sqrt{%
\left| \mathcal{K}\right| }}. 
\]

\noindent This means that the canonical coordinate $\sigma $ has been
transformed according to $\sigma \longrightarrow \sigma \sqrt{\left| 
\mathcal{K}\right| }$, and in order to retain the hamiltonian structure of
the microscopic equations of motion the $\sigma $ - component of the force
should also be transformed as $\widehat{F}_\sigma \longrightarrow
-sign\left( \mathcal{K}\right) \widehat{F}_\sigma $. Taking into account the
fact that the Balescu-Lenard collision integral is proportional to the
square of the fluctuating force we can write

\[
{\mathcal{J}}_{col}^{\left( BL\right) }\left( \widehat{\mathbf{x}},\widehat{%
\mathbf{p}}^{\left( k\right) };\theta \right) =\frac{\pi q^4nR}{\varepsilon
_0^2\beta _o^4\gamma _o^4E_o^2}\widehat{\mathbf{\nabla }}_p\cdot \int \frac{%
d^3{\mathbf{k}}d^3\widehat{\mathbf{p}}_1^{\left( k\right) }}{\left( 2\pi
\right) ^3}\delta \left( {\mathbf{k}}\cdot \widetilde{\mathbf{p}}^{\left(
k\right) }-{\mathbf{k}}\cdot \widetilde{\mathbf{p}}_1^{\left( k\right)
}\right) * 
\]

\[
\ast \frac{{\mathbf{kk}}}{\left| \widetilde{\epsilon }\left( {R\mathbf{k}}\cdot %
\widetilde{\mathbf{p}}^{\left( k\right) },\mathbf{k}\right) \right| ^2\left(
k_x^2+k_z^2+k_\sigma ^2/\gamma _o^2\left| {\mathcal{K}}\right| \right) ^2}%
\cdot 
\]

$$
\cdot \left[ f\left( {\mathbf{x}},\widehat{\mathbf{p}}_1^{\left( k\right)
};\theta \right) \widehat{\mathbf{\nabla }}_pf\left( \widehat{\mathbf{x}}%
,\widehat{\mathbf{p}}^{\left( k\right) };\theta \right) -f\left( 
{\mathbf{x}},\widehat{\mathbf{p}}^{\left( k\right) };\theta \right) \widehat{%
\mathbf{\nabla }}_{p_1}f\left( \widehat{\mathbf{x}},\widehat{%
\mathbf{p}}_1^{\left( k\right) };\theta \right) \right] ,  \eqno (5.6)
$$

\noindent where

\[
\widetilde{\mathbf{p}}^{\left( k\right) }=\left( \widetilde{p}_x^{\left(
k\right) },\widetilde{p}_z^{\left( k\right) },\widehat{\eta }^{\left(
k\right) }\right) . 
\]

\noindent Handling the integral

$$
\widehat{\mathbf{I}}_L\left( {\mathbf{g}}\right) =\int d^3{\mathbf{k}}\frac{%
{\mathbf{kk}}}{\left( k_x^2+k_z^2+k_\sigma ^2/\gamma _o^2\left| {\mathcal{K}}%
\right| \right) ^2}\delta \left( {\mathbf{k\cdot g}}\right) ,\qquad \quad
\left( {\mathbf{g}}=\widetilde{\mathbf{p}}^{\left( k\right) }-\widetilde{%
\mathbf{p}}_1^{\left( k\right) }\right)  \eqno (5.7)
$$

\noindent by choosing a reference frame in which the vector $\mathbf{g}$
points along the $\sigma $ - axis, and using cylindrical coordinates in this
frame we find

\[
\widehat{\mathbf{I}}_L\left( {\mathbf{g}}\right) =\int\limits_0^\infty
dk_{\perp }k_{\perp }\int\limits_0^{2\pi }d\Phi \int\limits_{-\infty
}^\infty dk_\sigma \delta \left( k_\sigma g\right) \frac 1{\left( k_{\perp
}^2+k_\sigma ^2/\gamma _o^2\left| {\mathcal{K}}\right| \right) ^2}* 
\]

$$
\ast \left( 
\begin{array}{c}
k_{\perp }\cos \Phi \\ 
k_{\perp }\sin \Phi \\ 
k_\sigma
\end{array}
\right) \left( 
\begin{array}{c}
k_{\perp }\cos \Phi \\ 
k_{\perp }\sin \Phi \\ 
k_\sigma
\end{array}
\right) =\frac \pi g\left( \widehat{\mathbf{I}}-{\mathbf{e}}_s{\mathbf{e}}%
_s\right) \int\limits_{k_D}^{k_L}\frac{dk_{\perp }}{k_{\perp }}=\frac{\pi 
{\mathcal{L}}}g\left( \widehat{\mathbf{I}}-{\mathbf{e}}_s{\mathbf{e}}_s\right) . 
\eqno (5.8)
$$

\noindent As was mentioned above in order to avoid logarithmic divergences
at both limits of integration on $k_{\perp }$ in (5.8) we have altered them
according to

$$
k_D=\frac 1{\gamma _or_D}\qquad ;\qquad k_L=\frac{4\pi \varepsilon _0k_BT}{%
\gamma _oq^2}.  \eqno (5.9)
$$

\noindent Thus the Coulomb logarithm ${\mathcal{L}}$ is defined as

$$
{\mathcal{L}}=\ln \left[ \frac{4\pi }{q^3\sqrt{n}}\left( \varepsilon
_0k_BT\right) ^{3/2}\right] .  \eqno (5.10)
$$

\noindent The tensor $\widehat{\mathbf{I}}_L\left( {\mathbf{g}}\right) $ can
be evaluated in an arbitrary reference frame to give

$$
\widehat{\mathbf{I}}_L\left( {\mathbf{g}}\right) =\frac{\pi {\mathcal{L}}}%
g\left( \widehat{\mathbf{I}}-\frac{{\mathbf{gg}}}{g^2}\right) .  \eqno (5.11)
$$

\noindent Finally the collision integral (5.6) can be represented in the
form of Landau as

\[
{\mathcal{J}}_{col}^{\left( L\right) }\left( \widehat{\mathbf{x}},\widehat{%
\mathbf{p}}^{\left( k\right) };\theta \right) =\frac{q^4nR\mathcal{L}}{8\pi
\varepsilon _0^2\beta _o^4\gamma _o^4E_o^2}\widehat{\mathbf{\nabla }}_p\cdot
\int d^3\widehat{\mathbf{p}}_1^{\left( k\right) }\widehat{\mathbf{G}}%
_L\left( {\mathbf{g}}\right) \cdot 
\]

$$
\cdot \left[ f\left( {\mathbf{x}},\widehat{\mathbf{p}}_1^{\left( k\right)
};\theta \right) \widehat{\mathbf{\nabla }}_pf\left( \widehat{\mathbf{x}}%
,\widehat{\mathbf{p}}^{\left( k\right) };\theta \right) -f\left( 
{\mathbf{x}},\widehat{\mathbf{p}}^{\left( k\right) };\theta \right) \widehat{%
\mathbf{\nabla }}_{p_1}f\left( \widehat{\mathbf{x}},\widehat{%
\mathbf{p}}_1^{\left( k\right) };\theta \right) \right] ,  \eqno (5.12)
$$

\noindent where

$$
\widehat{\mathbf{G}}_L\left( \mathbf{g}\right) =\frac 1g\left( \widehat{%
\mathbf{I}}-\frac{{\mathbf{gg}}}{g^2}\right)  \eqno (5.13)
$$

\noindent is the Landau tensor \cite{Balescu}.

\section{The Local Equilibrium State and Approximate Collision Integral.}

The local equilibrium state is defined as a solution to the equation

$$
{\mathcal{J}}_{col}\left( \widehat{\mathbf{x}},\widehat{\mathbf{p}}^{\left(
k\right) };\theta \right) =0,  \eqno (6.1)
$$

\noindent where the collision integral is taken either in Balescu-Lenard or
Landau form. This solution is well-known to be the Maxwellian distribution

$$
f_q\left( \widehat{\mathbf{x}},\widehat{\mathbf{p}}^{\left( k\right)
};\theta \right) =\rho \left( \frac R{2\pi \epsilon }\right) ^{3/2}\exp
\left[ -\frac R{2\epsilon }\left( \widehat{\mathbf{p}}^{\left( k\right) }-%
\mathbf{u}\right) ^2\right] ,  \eqno (6.2a)
$$

$$
\int d^3\widehat{\mathbf{x}}\rho \left( \widehat{\mathbf{x}};\theta \right)
=V,  \eqno (6.2b)
$$

\noindent where $\rho \left( \widehat{\mathbf{x}};\theta \right) $, $%
\epsilon \left( \widehat{\mathbf{x}};\theta \right) $ and ${\mathbf{u}}\left( 
\widehat{\mathbf{x}};\theta \right) $ are functions of $\widehat{\mathbf{x}}$
and $\theta $. It should be clear that the local equilibrium state is not a
true thermodynamic equilibrium state, since the latter must be homogeneous
and stationary. To prove that the distribution (6.2) is a solution of (6.1)
when the collision integral is taken in Landau form (5.12) it is sufficient
to take into account the obvious identity

$$
\widehat{\mathbf{G}}_L\left( {\mathbf{a}}\right) \cdot {\mathbf{a}} = % 
{\mathbf{a}}^T\cdot % 
\widehat{\mathbf{G}}_L\left( \mathbf{a}\right) =0.  \eqno (6.3)
$$

\noindent Next we note that the Landau collision integral (5.12) can be
written as a nonlinear Fokker-Planck operator

$$
{\mathcal{J}}_{col}^{\left( L\right) }\left( \widehat{\mathbf{x}},\widehat{%
\mathbf{p}}^{\left( k\right) };\theta \right) ={\mathcal{B}}\left[ \widehat{%
\mathbf{\nabla }}_p\cdot \left( \widehat{\mathcal{D}}\cdot \widehat{\mathbf{%
\nabla }}_p\right) -\widehat{\mathbf{\nabla }}_p\cdot {\mathbf{A}}\right]
f\left( \widehat{\mathbf{x}},\widehat{\mathbf{p}}^{\left( k\right) };\theta
\right) ,  \eqno (6.4)
$$

\noindent where

$$
{\mathcal{B}}=\frac{\mathcal{L}}{8\pi r_D^4\gamma _o^4}\frac{\epsilon ^2}{nR},
\eqno (6.5a)
$$

$$
\widehat{\mathcal{D}}=\int d^3\widehat{\mathbf{p}}_1^{\left( k\right) }%
\widehat{\mathbf{G}}_L\left( \mathbf{g}\right) f\left( \widehat{\mathbf{p}}%
_1^{\left( k\right) }\right) \qquad ;\qquad {\mathbf{A}} = \int d^3\widehat{%
\mathbf{p}}_1^{\left( k\right) }\widehat{\mathbf{G}}_L\left( {\mathbf{g}}%
\right) \cdot \widehat{\mathbf{\nabla }}_{p_1}f\left( \widehat{\mathbf{p}}%
_1^{\left( k\right) }\right) .  \eqno (6.5b)
$$

\noindent Our goal in what follows will be to match the transition to the
unified kinetic equation. To approach this it is sufficient to compute the
drift and diffusion coefficients (6.5b) using the local equilibrium
distribution (6.2). A more systematic approximation methods using the
linearized Landau collision integral can be found in \cite{Balescu}. Going
over to the new variable

$$
{\mathbf{C}}=\sqrt{\frac R{2\epsilon }}\left( \widehat{\mathbf{p}}^{\left(
k\right) }-{\mathbf{u}}\right) =\sqrt{\frac R{2\epsilon }}\delta \widehat{%
\mathbf{p}}  \eqno (6.6)
$$

\noindent we write

$$
\widehat{\mathcal{D}}\left( \mathbf{C}\right) =\frac{2\epsilon }R\int d^3%
{\mathbf{C}}_1\widehat{\mathbf{G}}_L\left( \mathbf{g}\right) f_q\left( {\mathbf{%
C}}_1\right) ={\mathcal{A}}_0\left( C\right) \widehat{\mathbf{G}}_L\left(
\delta \widehat{\mathbf{p}}\right) +{\mathcal{A}}_1\left( C\right) \frac{%
\delta \widehat{\mathbf{p}}\delta \widehat{\mathbf{p}}}{\left( \delta 
\widehat{p}\right) ^4},  \eqno (6.7)
$$

\noindent where ${\mathbf{g=C-C}}_1$ and ${\mathcal{A}}_0$, ${\mathcal{A}}_1$ are
functions of the modulus of the vector ${\mathbf{C}}$

$$
{\mathcal{A}}_1\left( C\right) =\delta \widehat{\mathbf{p}}\cdot \widehat{%
\mathcal{D}}\cdot \delta \widehat{\mathbf{p}}=\frac{2\epsilon }R\int d^3%
{\mathbf{C}}_1\delta \widehat{\mathbf{p}}\cdot \widehat{\mathbf{G}}_L\left( 
{\mathbf{g}}\right) \cdot \delta \widehat{\mathbf{p}}f_q\left( {\mathbf{C}}%
_1\right) ,  \eqno (6.8a)
$$

$$
{\mathcal{A}}_0\left( C\right) =\frac{\delta \widehat{p}}2Sp\left( \widehat{%
\mathcal{D}}\right) -\frac 1{2\delta \widehat{p}}{\mathcal{A}}_1\left(
C\right) ,  \eqno (6.8b)
$$

$$
Sp\left( \widehat{\mathcal{D}}\right) =\frac{4\epsilon }R\int d^3{\mathbf{C}}_1%
\frac{f_q\left( {\mathbf{C}}_1\right) }{\left| {\mathbf{C-C}}_1\right| }. 
\eqno (6.8c)
$$

\noindent To compute the integrals (6.8a) and (6.8c) we use spherical
coordinates in a reference frame in which vector $\mathbf{C}$ points along
the $\sigma $ - axis. We find

\[
Sp\left( \widehat{\mathcal{D}}\right) =\frac{2\rho }{\pi ^{3/2}}\sqrt{\frac
R{2\epsilon }}\int\limits_0^\infty dC_1C_1^2\int\limits_0^{2\pi }d\Phi
\int\limits_{-1}^1d\cos \Theta \frac{e^{-C_1^2}}g, 
\]

\[
{\mathcal{A}}_1\left( C\right) =\frac \rho {\pi ^{3/2}}\sqrt{\frac{2\epsilon }R%
}\int\limits_0^\infty dC_1C_1^2\int\limits_0^{2\pi }d\Phi
\int\limits_{-1}^1d\cos \Theta \frac{e^{-C_1^2}}g\left[ C^2-\frac{\left(
C^2-CC_1\cos \Theta \right) ^2}{g^2}\right] , 
\]

\noindent where we have used ${\mathbf{g\cdot C}}=C^2-CC_1\cos \Theta $.
Changing variables in the above integrals according to

\[
g^2=C^2+C_1^2-2CC_1\cos \Theta \qquad ;\qquad d\cos \Theta =-\frac g{CC_1}dg 
\]

\noindent yields the result:

\[
Sp\left( \widehat{\mathcal{D}}\right) =\frac{4\rho }C\sqrt{\frac R{2\pi
\epsilon }}\int\limits_0^\infty dC_1C_1e^{-C_1^2}\int\limits_{\left|
C-C_1\right| }^{C+C_1}dg= 
\]

\[
=\frac{8\rho }C\sqrt{\frac R{2\pi \epsilon }}\left[
\int\limits_0^CdC_1C_1^2e^{-C_1^2}+C\int\limits_C^\infty
dC_1C_1e^{-C_1^2}\right] =\frac{2\rho }C\sqrt{\frac R{2\epsilon }}{\rm{%
erf}}\left( C\right) . 
\]

\noindent and similarly

\[
{\mathcal{A}}_1\left( C\right) =\frac{2\rho }{C\sqrt{\pi }}\sqrt{\frac{%
2\epsilon }R}\int\limits_0^\infty dC_1C_1e^{-C_1^2}\int\limits_{\left|
C-C_1\right| }^{C+C_1}dg\left[ C^2-\frac{\left( g^2-C_1^2+C^2\right) ^2}{4g^2%
}\right] = 
\]

\[
=\frac{8\rho }{3C\sqrt{\pi }}\sqrt{\frac{2\epsilon }R}\left[
\int\limits_0^CdC_1C_1^4e^{-C_1^2}+C^3\int\limits_C^\infty
dC_1C_1e^{-C_1^2}\right] =\frac \rho C\sqrt{\frac{2\epsilon }R}\left[
1-C\frac d{dC}\right] {\rm{erf}}\left( C\right) , 
\]

\noindent where ${\rm{erf}}\left( C\right) $ is the error function. Thus
for the coefficients ${\mathcal{A}}_0$ and ${\mathcal{A}}_1$ in (6.8) we have

$$
{\mathcal{A}}_0\left( C\right) =\rho {\mathcal{C}}\left( C\right) \qquad ;\qquad 
{\mathcal{A}}_1\left( C\right) =\frac \rho C\sqrt{\frac{2\epsilon }R}\left[
1-C\frac d{dC}\right] {\rm{erf}}\left( C\right) ,  \eqno (6.9)
$$

\noindent where

$$
{\mathcal{C}}\left( C\right) =\frac 1{2C^2}\left( 2C^2-1+C\frac d{dC}\right) 
{\rm{erf}}\left( C\right)  \eqno (6.10)
$$

\noindent is the Chandrasekhar function. The drift vector can be written as

$$
{\mathbf{A}}\left( \delta \widehat{\mathbf{p}}\right) =-\frac R\epsilon 
\widehat{\mathcal{D}}\left( \delta \widehat{\mathbf{p}}\right) \cdot \delta 
\widehat{\mathbf{p}}=-\frac R\epsilon {\mathcal{A}}_1\left( C\right) \frac{%
\delta \widehat{\mathbf{p}}}{\left( \delta \widehat{p}\right) ^2}. 
\eqno (6.11)
$$

\noindent The drift and diffusion coefficients can be further evaluated by
substituting $\delta \widehat{\mathbf{p}}$ with the r.m.s. value

$$
\left( \delta \widehat{p}_i\right) _{rms}=\sqrt{\frac{\epsilon \left( 
\widehat{\mathbf{x}};\theta \right) }R}\qquad ;\qquad \left( C_i\right)
_{rms}=\frac 1{\sqrt{2}}.  \eqno (6.12)
$$

\noindent Thus we obtain

$$
\widehat{\mathcal{D}}=D\widehat{\mathbf{I}}\qquad ;\qquad {\mathbf{A}}=-\frac
R\epsilon D\delta \widehat{\mathbf{p}},  \eqno (6.13)
$$

\noindent where

$$
D=\frac 13Sp\left( \widehat{\mathcal{D}}\right) =\frac{2R^{3/2}{\rm{erf}}%
\left( \sqrt{3/2}\right) }{\left( 3\epsilon \right) ^{3/2}}\frac{\epsilon
\left( \widehat{\mathbf{x}};\theta \right) }R  \eqno (6.14)
$$

\noindent for $\rho \left( \widehat{\mathbf{x}};\theta \right) \sim 1$. This
enables us to cast (6.4) into the form

$$
{\mathcal{J}}_{col}^{\left( L\right) }\left( \widehat{\mathbf{x}},\widehat{%
\mathbf{p}}^{\left( k\right) };\theta \right) =\frac 1{\theta _{rel}}\left\{ 
\frac{\epsilon \left( \widehat{\mathbf{x}};\theta \right) }R\widehat{\mathbf{%
\nabla }}_p^2+\widehat{\mathbf{\nabla }}_p\cdot \left[ \widehat{\mathbf{p}}%
^{\left( k\right) }-{\mathbf{u}}\left( \widehat{\mathbf{x}};\theta \right)
\right] \right\} f\left( \widehat{\mathbf{x}},\widehat{\mathbf{p}}^{\left(
k\right) };\theta \right) ,  \eqno (6.15)
$$

\noindent where

$$
\theta _{rel}=\frac{12\sqrt{3}\pi }{{\rm{erf}}\left( \sqrt{3/2}\right) }%
\frac{nr_D^4\gamma _o^4}{{\mathcal{L}}\sqrt{\epsilon R}}  \eqno (6.16)
$$

\noindent is the relaxation ''time''.

\section{The Generalized Kinetic Equation.}

The transition to local equilibrium, that is the kinetic stage of
relaxation, is described by the Balescu-Lenard or the Landau kinetic
equation. The latter with due account of the approximate collision integral
(6.15) can be written as

$$
\frac{\partial f}{\partial \theta }+R\left( \widehat{\mathbf{p}}^{\left(
k\right) }\cdot \widehat{\mathbf{\nabla }}_x\right) f+R\left( \widehat{%
\mathbf{F}}\cdot \widehat{\mathbf{\nabla }}_p\right) f=\frac 1{\theta
_{rel}}\left[ \frac \epsilon R\widehat{\mathbf{\nabla }}_p^2+\widehat{%
\mathbf{\nabla }}_p\cdot \left( \widehat{\mathbf{p}}^{\left( k\right) }-%
\mathbf{u}\right) \right] f,  \eqno (7.1)
$$

\[
\widehat{\mathbf{F}}=\widehat{\mathbf{F}}_0+\left\langle \widehat{\mathbf{F}}%
\right\rangle . 
\]

\noindent It is well-known \cite{Gardiner} that the kinetic equation (7.1)
is equivalent to the system of Langevin equations:

$$
\frac{d\widehat{\mathbf{x}}}{d\theta }=R\widehat{\mathbf{p}}^{\left(
k\right) }\quad ;\quad \frac{d\widehat{\mathbf{p}}^{\left( k\right) }}{%
d\theta }=-\frac 1{\theta _{rel}}\left( \widehat{\mathbf{p}}^{\left(
k\right) }-{\mathbf{u}}\right) +R\widehat{\mathbf{F}}+\sqrt{\frac \epsilon
{R\theta _{rel}}}{\vec {\xi }}\left( \theta \right) ,  \eqno (7.2)
$$

\noindent where ${\vec {\xi }}\left( \theta \right) $ is a white-noise
random variable with formal correlation properties

$$
\left\langle {\vec {\xi }}\left( \theta \right) \right\rangle =0\qquad
;\qquad \left\langle \xi _m\left( \theta \right) \xi _n\left( \theta
_1\right) \right\rangle =2\delta _{mn}\delta \left( \theta -\theta _1\right)
.  \eqno (7.3)
$$

\noindent The generalized kinetic equation (2.7) describes the evolution of
the beam for time scales greater than the relaxation time $\theta _{rel}$.
In order to determine the additional collision integral $\widetilde{\mathcal{%
J}}\left( \widehat{\mathbf{x}},\widehat{\mathbf{p}}^{\left( k\right)
};\theta \right) $ we use the method of adiabatic elimination of fast
variables, which in our case are the kinetic momenta $\widehat{\mathbf{p}}%
^{\left( k\right) }$. In the limit of small times $\theta _{rel}$ (compared
to the time scale of physical interest) the second equation (7.2) relaxes
sufficiently fast to the quasi-stationary (local equilibrium) state for
which $d\widehat{\mathbf{p}}^{\left( k\right) }/d\theta \longrightarrow 0$.
Thus we find

$$
\widehat{\mathbf{p}}^{\left( k\right) }={\mathbf{u}}+R\theta _{rel}\widehat{%
\mathbf{F}}+\sqrt{\frac{\epsilon \theta _{rel}}R}{\vec {\xi }}\left( \theta
\right)  \eqno (7.4)
$$

\noindent and substituting this into the first of equations (7.2) we arrive
at

$$
\frac{d\widehat{\mathbf{x}}}{d\theta }=R{\mathbf{u}}+R^2\theta _{rel}\widehat{%
\mathbf{F}}+\sqrt{\epsilon R\theta _{rel}}{\vec {\xi }}\left( \theta \right)
.  \eqno (7.5)
$$

\noindent The above equation (7.5) governs the evolution of particles within
the elementary cell of continuous medium, where local equilibrium state is
established. Such a coarse-graining procedure gives rise to the additional
collision integral in the generalized kinetic equation (2.7). The latter
follows straighforwardly from (7.5) and can be written in the form:

$$
\widetilde{\mathcal{J}}\left( \widehat{\mathbf{x}},\widehat{\mathbf{p}}%
^{\left( k\right) };\theta \right) =R\theta _{rel}\left\{ \widehat{\mathbf{%
\nabla }}_x\cdot \left[ \epsilon \left( \widehat{\mathbf{x}};\theta \right) 
\widehat{\mathbf{\nabla }}_x\right] -R\left( \widehat{\mathbf{\nabla }}%
_x\cdot \widehat{\mathbf{F}}\right) \right\} f\left( \widehat{\mathbf{x}},%
\widehat{\mathbf{p}}^{\left( k\right) };\theta \right) .  \eqno (7.6)
$$

\section{Concluding Remarks.}

In the present paper we have studied the role of electromagnetic
interactions between particles on the evolution of a high energy beam. The
interparticle forces we have considered here are due to space charge alone. 
Starting with the reversible dynamics of individual particles and applying a
smoothing procedure over the physically infinitesimal spacial scales, we
have derived a generalized kinetic equation for kinetic, hydrodynamic and
diffusion processes. 

We would like to point out an important feature of the approach presented 
in this work. The irreversibility of beam evolution is introduced at the 
very beginning in the initial equation (2.1) for the microscopic phase 
space density. Smoothing destroys information about the motion of 
individual particles within the unit cell of continuous medium, hence 
the reversible description becomes no longer feasible. Details of 
particle dynamics become lost and motion smears out due to dynamic 
instability, and to the resulting mixing of trajectories in phase space.

The collision integral for a high energy beam has been derived (Sections IV
and V) in the form of Balescu-Lenard and Landau. This collision term scales
as $E_o^{-6}$ ($E_o$ is the energy of the synchronous particle) which
comprises a negligibly weak dissipative mechanism for high energy beams.

To accomplish the transition to the generalized kinetic equation the Landau
collision term has been simplified by linearizing it around the local
equilibrium distribution. The latter suggests a close relation between
equilibrium beam emittance and the temperature of the beam.

Finally in Section VII we have derived the additional dissipative term due
to the redistribution of particle coordinates. This has been achieved by
applying the method of adiabatic elimination of fast variables (kinetic
momenta). The physical grounds for this application is provided the fact that
within the physically infinitesimal confinement the relatively slow process
of smear in configuration space is induced by the sufficiently fast
relaxation of particle velocities towards a local equilibrium state. It
maybe worthwhile to note that a more systematic approach involving the
projection operator technique \cite{Gardiner} could be used to derive the
additional collision integral in the generalized kinetic equation.

\section{Acknowledgements.}

It is a pleasure to thank Pat Colestock, Jim Ellison and Alejandro Aceves 
for helpful discussions on the subject touched upon in the present paper, 
as well as David Finley and Steve Holmes for their support of this work.

\end{document}